\documentclass[aps,pre,preprint,groupedaddress]{revtex4-1}

\usepackage{graphicx}
\usepackage{dcolumn}
\usepackage{amsmath}    
\usepackage{amssymb}
\usepackage{bm} 
\usepackage{hyperref}
\usepackage{latexsym}
\usepackage{verbatim}
\usepackage{color}
\usepackage{subfigure}
\def\beq{\begin{equation}}
\def\eeq{\end{equation}}

\def\etal{{\it et al.}}

\def\Q{\mbox{\sffamily\bfseries Q}}

\newcommand{\bsf}[1]{\textsf{\textbf{#1}}}

\def\sigtens{\mbox{\boldmath $\sigma$\unboldmath}}

\def\beq{\begin{equation}}                           
\def\eeq{\end{equation}}                           
\def\bea{\begin{eqnarray}}                           
\def\eea{\end{eqnarray}}        

\begin{document}


\title{Aspects of the density field in an active nematic}


\author{Shradha Mishra}
\email[]{shradha.mishra@bose.res.in}
\affiliation{Department of Theoretical Sciences, S N Bose National Centre for
Basic Sciences, Kolkata 700098}
\author{Sanjay Puri}
\email[]{puri@mail.jnu.ac.in}
\affiliation{School of Physical Sciences, Jawaharlal Nehru University, New
Delhi 110 067}
\author{Sriram Ramaswamy}
\email[]{sriram@tifrh.res.in}
\affiliation{TIFR Centre for Interdisciplinary Sciences, Hyderabad 500 075,
India}
\altaffiliation{On leave from the Department of Physics, Indian Institute
of Science, Bangalore}

\date{\today}

\begin{abstract}
Active nematics are conceptually the simplest orientationally ordered phase of
self-driven particles, but have proved to be a perennial source of surprises.
We show here through numerical solution of coarse-grained equations for
order parameter and density that the growth of the active nematic phase from the
isotropic phase is necessarily accompanied by a clumping of the density. The
growth kinetics of the density domains is shown to be faster than the 1/3-law
expected for variables governed by a conservation law. Other results presented
include the suppression of density fluctuations in the stationary ordered
nematic by the imposition of an orienting field. We close by posing some
open questions. 
\end{abstract}
\maketitle

\section{Introduction}  \label{sec:intro}
\subsection{Background: nematic and polar order, active and passive}
\label{sub:nemactpass}
The nematic \cite{pgdg,chandrasekhar} is the simplest
liquid-crystalline phase,
a true fluid in all directions, but with a preferred axis. It possesses
uniaxial, fore-aft symmetric orientational order and no positional order. This
paper is concerned with systems with this spatial symmetry, but without
time-reversal symmetry: each constituent particle of the system is endowed
with the ability to consume free energy which it transduces into movement
\cite{schweitzer,epjst}. This ordered phase of self-driven matter
\cite{tonertusr,annurev,rmp}
is known as an active nematic \cite{sradititoner}.  

Early interest \cite{vicsek,tonertu,vicsekreview} in flocks as ordered
phases of nonequilibrium condensed matter focused naturally on the case where
the
individual constituents and the emergent ordered state observe the distinction
between
ahead and behind. A flock, after all, is a collection of creatures that is going
somewhere. The idea of classic liquid-crystalline order in living matter, with
anisotropy but no polarity, appeared initially in analyses of patterns in
aggregates of elongated cells \cite{gruler}, but the idea that such phases were
wildly different from their dead thermal-equilibrium counterparts emerged only
gradually \cite{sradititoner,tonertusr}. The subject has grown and evolved
considerably
since, with studies on the nature of density fluctuations
\cite{vjmenonsr,chateginellimontagne,shradhasr}, the passage from microscopic to
coarse-grained descriptions \cite{shradhathesis,shi1,selfreg} and the discovery
and elucidation of intrinsic instabilities in the latter \cite{shi1,njop},
and the unique
properties of topological defects in active nematics 
\cite{vjmenonsr,dogic,giomi2,thampi,pismen,shi2}. Clear numerical evidence
is now available for the existence of a statistically homogeneous
state with nematic order in an noisy active system, in a study \cite{ngo2014}
that poses new puzzles regarding the scaling properties of density
fluctuations. 

To set the stage, we review briefly some basic findings from the initial
literature on active nematics. Throughout this paper we will consider systems in
which the total number of particles is the only conserved quantity. We will
assume that all parts of the system under consideration are homogeneously in
contact with a passive momentum sink, so that the velocity field of the system
relaxes on a finite timescale to a value determined by the slow variables,
\textit{viz.}, the number density field $\rho$ and the traceless, symmetric,
second-rank tensor $\bsf{Q}$ characterising nematic order
\cite{pgdg,chandrasekhar}.
Effects relating to hydrodynamic flow \cite{rmp,dogic,giomi2,thampi,pismen} will
be mentioned in passing, if at all. 

\subsection{Currents from orientational distortions} \label{sub:curvcurr}
The first hint of peculiarities in the dynamics and statistics of $\rho$ in an
active nematic came from noting \cite{sradititoner} that on general grounds
the particle current $\mathbf{J} = \rho {\bf v}$, which defines the velocity
field ${\bf v}$, is permitted to have a nonequilibrium contribution
\begin{equation}
\label{jact}
\mathbf{J}_{act} \propto \rho \nabla \cdot \bsf{Q} 
\end{equation}
proportional to the orientational curvature. In more detail, following
refs. \cite{sradititoner,annurev,rmp}, the origin of
the term can be understood as follows. Consider a general collection of
particles, interacting with each other and moving while in contact with a
dissipative substrate that serves as a sink for momentum. Balancing friction
with the substrate against other force densities ${\bf \mathcal{F}}$ arising
from interparticle interactions or external driving fields yields ${\bf v} =
\bsf{M}\cdot{\bf \mathcal{F}}$ where $\bsf{M}$ is a mobility -- the inverse of a
damping coefficient. For the case where the non-frictional forces come entirely
from interparticle interactions, and are therefore momentum conserving, we can
write ${\bf \mathcal{F}} = -\nabla \cdot \sigtens$ where $\sigtens$ is the
stress tensor. The deviatoric stress for an \textit{equilibrium} nematic system
with a free-energy functional $F$, to leading order in $\bsf{Q}$, is
proportional to $\delta F / \delta \bsf{Q}$, and thus vanishes for the mean
state of nematic order. Leading-order stresses for a perturbed nematic involve
gradients of the order parameter, weighted by the Frank elastic constants.
However, for a nematic steady state away from thermal equilibrium, the order
parameter value is not determined by free-energy minimisation, and a piece of
the stress proportional to $\bsf{Q}$ itself, rather than to its conjugate
field, is permitted, leading to the form \eqref{jact} at leading order in a
gradient expansion.

The resulting statistical steady state must then be governed by a dynamic
balance between two kinds of mass fluxes: $\mathbf{J}_{act}$ driven by
orientational inhomogeneities and a diffusional current $\propto \nabla \rho$
ironing out density inhomogeneities. Deep in the nematically ordered phase the
dominant contribution to $\nabla \cdot \bsf{Q}$ will come from spatial
variations in the principal axes of $\bsf{Q}$, not its magnitude. Fluctuations
in density should then be comparable in magnitude to those in the orientation.
The latter, in the nematic, which is a phase in which continuous rotation
invariance is broken spontaneously, should diverge at small wavenumber. So,
therefore, should fluctuations in the number density, by the foregoing argument
\cite{sradititoner}: regions in an active nematic containing on average $N$
particles should display fluctuations in the number with a standard deviation
growing more rapidly than $\sqrt{N}$. Giant number fluctuations have now been
seen in experimental \cite{vjmenonsr} and numerical realisations
\cite{chateginellimontagne} of active nematics, but our understanding of this
deceptively simple system continues to evolve \cite{ngo2014,putzig2014}.

\subsection{Microscopic model of apolar flocking} \label{sub:chatemodel}
We summarise briefly the microscopic model of Chat\'e, Ginelli and Montagne
\cite{chateginellimontagne} for active, uniaxial, headless particles with
a tendency to align with each other.
The particles, labelled by $\alpha$, 
have positions ${\bf R}_{\alpha}$ and orientations
$(\cos \theta_{\alpha}, \sin \theta_{\alpha})$ with respect to a fixed
reference frame. Particles move synchronously
at discrete times $t$ by a fixed distance $\epsilon$.
in the direction defined by $\theta_{\alpha}$ or $\theta_{\alpha}+\pi$ with
equal probability. The time step is taken as unity, and the updated
orientation is  defined in two stages: first, orient the particle
parallel to the mean of its neighbours, along the direction of the first
eigenvector of the average of the individual orientation tensors
\begin{equation}
\bsf{Q} _{\beta} = \left(\begin{array}{cc}
\cos^2\theta_{\beta} - \frac{1}{2} & \cos\theta_{\beta}
\sin\theta_{\beta}\\
\cos\theta_{\beta} \sin\theta_{\beta}  &
\sin^2\theta_{\beta}  -\frac{1}{2}\end{array} \right)
\label{dir}
\end{equation}
of all particles $\beta$ including $\alpha$ with $|{\bf R}_{\beta} - {\bf
R}_{\alpha}|$ less than a fixed interaction range, taken to be
substantially larger than the individual displacements $\epsilon$. Next, add to
it a small noise:
\begin{equation}
\theta_{\alpha}(t+1) \to  \theta_{\alpha}(t+1) + \sigma
\zeta_{\alpha}(t)
\label{thetaeq}
\end{equation}
where $\zeta_{\alpha}$
is distributed uniformly on $[−\pi, \pi]$, identically and independently for
each $\alpha$ and $t$, and $\sigma$ controls the noise amplitude.

Coarse-graining the above rules was shown \cite{shradhathesis,njop} to generate
the active current $\mathbf{J}_{act} \propto \rho \nabla \cdot
\bsf{Q}$ \cite{sradititoner} discussed above. The complete
PDEs for the number density
\begin{equation}
\rho({\bf r}, t) = \sum_{\alpha}\delta({\bf r}-{\bf R}_{\alpha}(t))
\label{rhodef}
\end{equation}
and the nematic tensor density
\begin{equation}
\bsf{w}({\bf r}, t)  \equiv \rho \bsf{Q}({\bf r}, t) ={
\sum_{\alpha}{\bsf{Q}_{\alpha}\delta({\bf r}-{\bf
R}_{\alpha}(t))} }
\label{qdef}
\end{equation}
were shown \cite{njop} to have the general form 
\begin{equation}
\frac{\partial \rho}{\partial t} = \frac{1}{2}\Delta \rho + \frac{1}{2}
{\bf \Gamma}:\bsf{w}
\label{densityequation}
\end{equation}
and
\begin{equation}
\frac{\partial \bsf{w}}{\partial t} = \mu \bsf{w} - 2 \xi \bsf{w}(\bsf{w} :
\bsf{w}) + \frac{1}{2}\Delta \bsf{w} + \frac{1}{8}{\bf \Gamma} \rho. 
\label{opequation}
\end{equation}
In (\ref{densityequation}) and (\ref{opequation}) $\Delta$ is the Laplacian, the coefficients $\mu$ and $\xi$
are calculable functions \cite{njop} of $\rho$ and the noise strength $\sigma$
in (\ref{thetaeq}).
For our numerical study we choose a simpler density-dependence of
$\mu$ and $\xi$ than calculated in \cite{njop}, as discussed in
section (\ref{sec:coarseclump}).  
The tensor differential operator ${\bf \Gamma}$ has components $\Gamma_{11}
= -\Gamma_{22} = \partial_1 \partial_1 - \partial_2 \partial_2$
and $\Gamma_{12} = \Gamma_{21} = 2 \partial_1 \partial_2$ so that ${\bf \Gamma}:
\bsf{w} = 2 \partial_{\alpha}\partial_{\beta} w_{\alpha \beta}$. 

\subsubsection{Multiplicative noise} \label{subsub:multnoise}
Equations (\ref{densityequation}) and (\ref{opequation}) represent only the
mean-field behaviour flowing from the microscopic  model . The
complete picture must include the noise sources induced in the coarse-grained
dynamics by the stochastic part of the microscopic dynamics, as discussed
in \cite{njop,shradhathesis}. The noise in (\ref{densityequation}) is the
divergence of a random current with covariance of order $\rho(\bsf{Q} +
\bsf{I}/2)$ where $\bsf{I}$ is the unit tensor, while that in
(\ref{opequation}), rewritten as an equation for $\bsf{Q}$, is of order
$1/\rho$. Crucially, $\rho$ and $\bsf{Q}$ in the foregoing refer to the
\textit{local, instantaneous} values of the fields in question. Such
density-dependent multiplicative noise is a generic consequence
\cite{dean} of the Central Limit Theorem together with the fact that $\rho$
and $\bsf{Q}$ are, respectively, the density of an extensive quantity and the
local expression of an intensive quantity. Stochastic-PDE treatments of dynamic
critical phenomena at equilibrium
\cite{HH} implicitly ignore such field dependence, presumably because it is 
irrelevant there. In active matter where density fluctuations are large across
an entire phase rather than just at phase boundaries, it could well have more
serious consequences \cite{njop}, but we will not explore this issue further in
this paper. In section(\ref{sec:coarseclump}) we will present results from a
numerical solution of (\ref{densityequation}) and (\ref{opequation}), mainly
without noise sources. 

\subsubsection{Proliferation of topological defects} \label{subsub:defprolif}
Topological defects in active nematics pose puzzles that are still to be
resolved -- and that, regrettably, we shall not resolve in this article. The
headless nature of nematic order means the defects have
half-integer strength, representing a rotation through $\pm \pi$ of the order
parameter upon circumambulating an elementary disclination \cite{pgdg}.
Structurally, defects of strength $-1/2$ have a 3-fold symmetric appearance
while strength $+1/2$ defects have a polarity and hence move in a directed
fashion because active nematics are out of equilibrium
\cite{vjmenonsr,giomi2,thampi,pismen}. In this paper we shall steer clear of
issues concerning the degradation of nematic order by defect proliferation. We
shall assume defects have a prohibitively high core energy and are thus not
produced at all in steady state on accessible length scales. In coarsening
studies, however, the onset of local order about the initial statistically
isotropic and homogeneous state forces the emergence of defects whose
annihilation leads to the growth of ordered domains.

\subsection{Problems of interest and summary of results} \label{subsec:results}
Here we present our results in brief, from a study of the noiseless
\cite{foot1noise} equations (\ref{densityequation}) and
(\ref{opequation}) describing the dynamics of the density and nematic order
parameter of an active nematic-forming system. (i) When quenched from
the isotropic to the nematic phase, nematically ordered domains grows with time
$t$, as expected. The growth law over the range explored is slower than
$t^{1/2}$, and is consistent with the behaviour
expected \cite{abray} for the growth of an equilibrium nematic. (ii) More
interestingly, the \textit{density} displays clumping and coarsening as well.
The associated length scale grows more rapidly than the $t^{1/3}$ normally
associated with domain growth for a conserved order parameter \cite{abray}. 
(iii) The domains for density and nematic order parameter both show a
characteristic three-armed morphology, with a topological defect of strength
$-1/2$ at the nodal point of the arms. (iv) Deep in the active nematic phase, we
impose an aligning field and measure numerically the steady-state density
fluctuations in the presence of a small additive noise. We find that for large
enough field the fluctuations saturate at large system size, at a value that
diverges as the field is taken to zero. The form of the dependence compares well
to analytical calculations from a linearised theory. The controlling role of the
field is a clear indicator of the underlying mechanism of giant density
fluctuations, namely, mass flux generated by orientational distortions. (v)
We note the strong departure from the standard
fluctuation-dissipation relation implied by the large density fluctuations:
the \textit{response} to a field coupling to the density is normal, the
\textit{fluctuations} of the density are anomalous. We also point out an
apparent incompatibility between the anisotropic density correlations
in the linearised theory of the active nematic \cite{sradititoner} and the
quasi-long-range character of two-dimensional nematic order.

The remainder of this paper is organised as follows: In section
\ref{sec:coarseclump} we present our study of the growth kinetics of the active
nematic from an isotropic background. In section
\ref{sec:extfield} we examine the effect of an aligning field on nematic and
density correlations. In section \ref{sec:otherissues} we discuss correlation
vs response, and the puzzle of anisotropic density correlations. We close with
a summary. 

\section{Coarsening and clumping} \label{sec:coarseclump}

\begin{figure}[htbp]
\begin{center}
\subfigure
{\includegraphics[height=9cm, width=10cm,
angle=0]{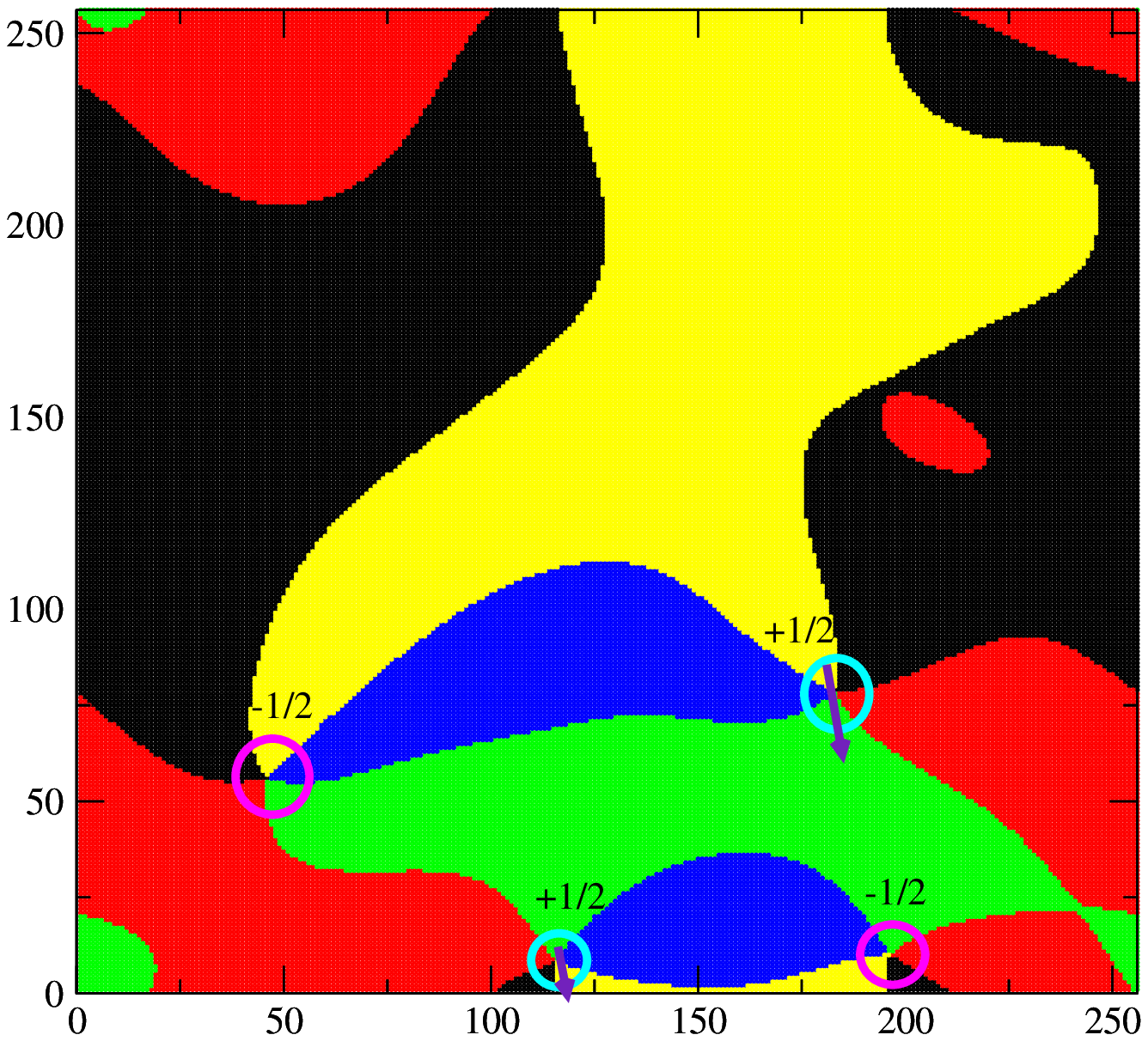}}
\subfigure{\includegraphics[height=9cm, width=10cm,
angle=0]{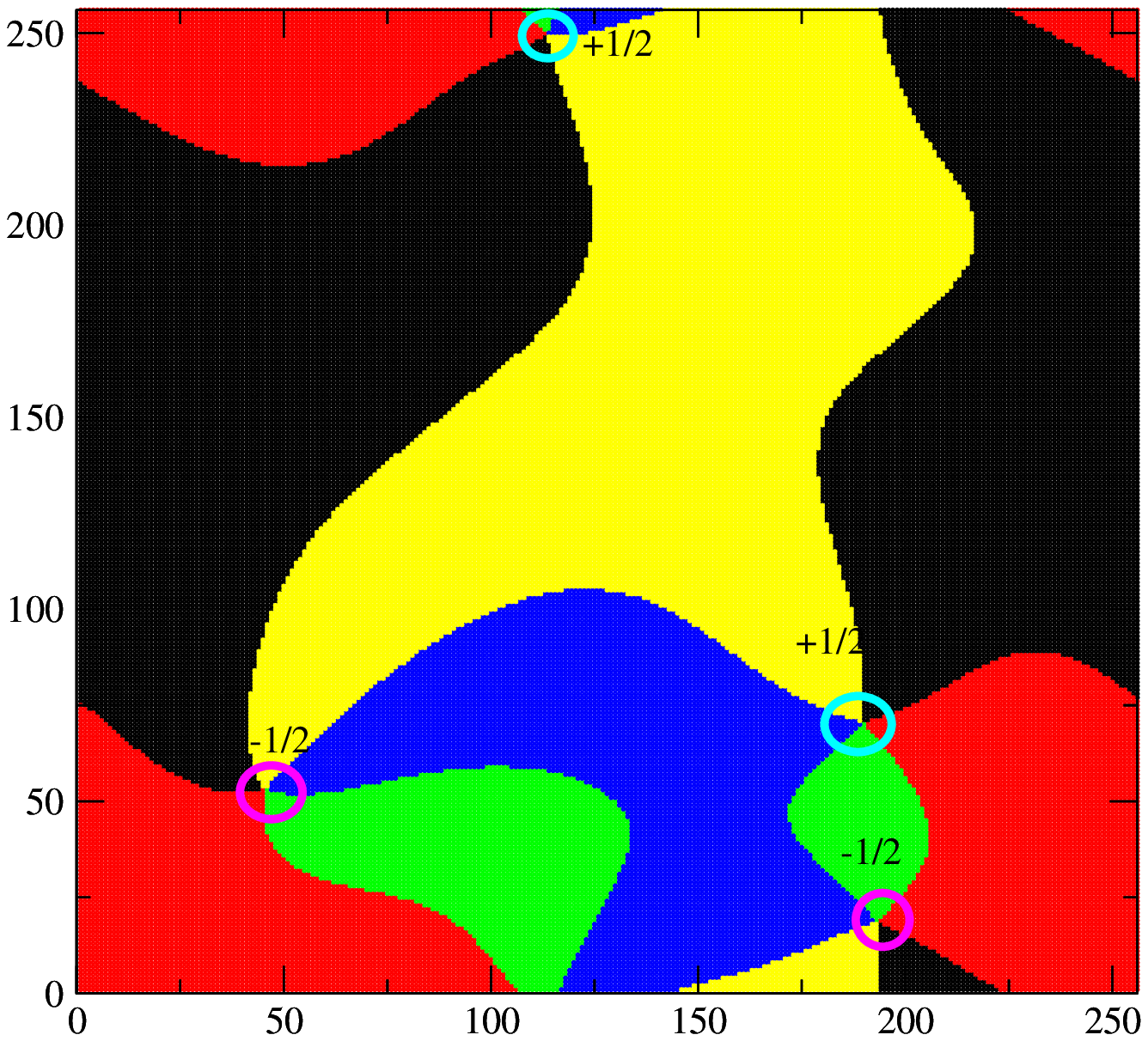}}
\caption{(color online) Two images of the angle field for nematic
order parameter during the coarsening process governed by Equations
(\ref{densityequation}) and (\ref{opequation}); the lower figure is at a
later time. Colours black-red-green-blue-yellow
denote angle ranges
$[0:0.2\pi]-[0.2\pi:0.4\pi]-[0.4\pi:0.6\pi]-[0.6\pi:0.8\pi]-[0.8\pi:\pi]$.
The meeting points of different colours are therefore point defects of strength
$\pm 1/2$. $+1/2$ and $-1/2$ defects are marked with red and green circles 
respectively. Arrows show the direction of motion of $+1/2$ defects.}
\label{defectpic}
\end{center}
\end{figure}

\begin{figure}[htbp]
\begin{center}
\subfigure(a){\includegraphics[height=6.0cm, width=6.0cm,
angle=270]{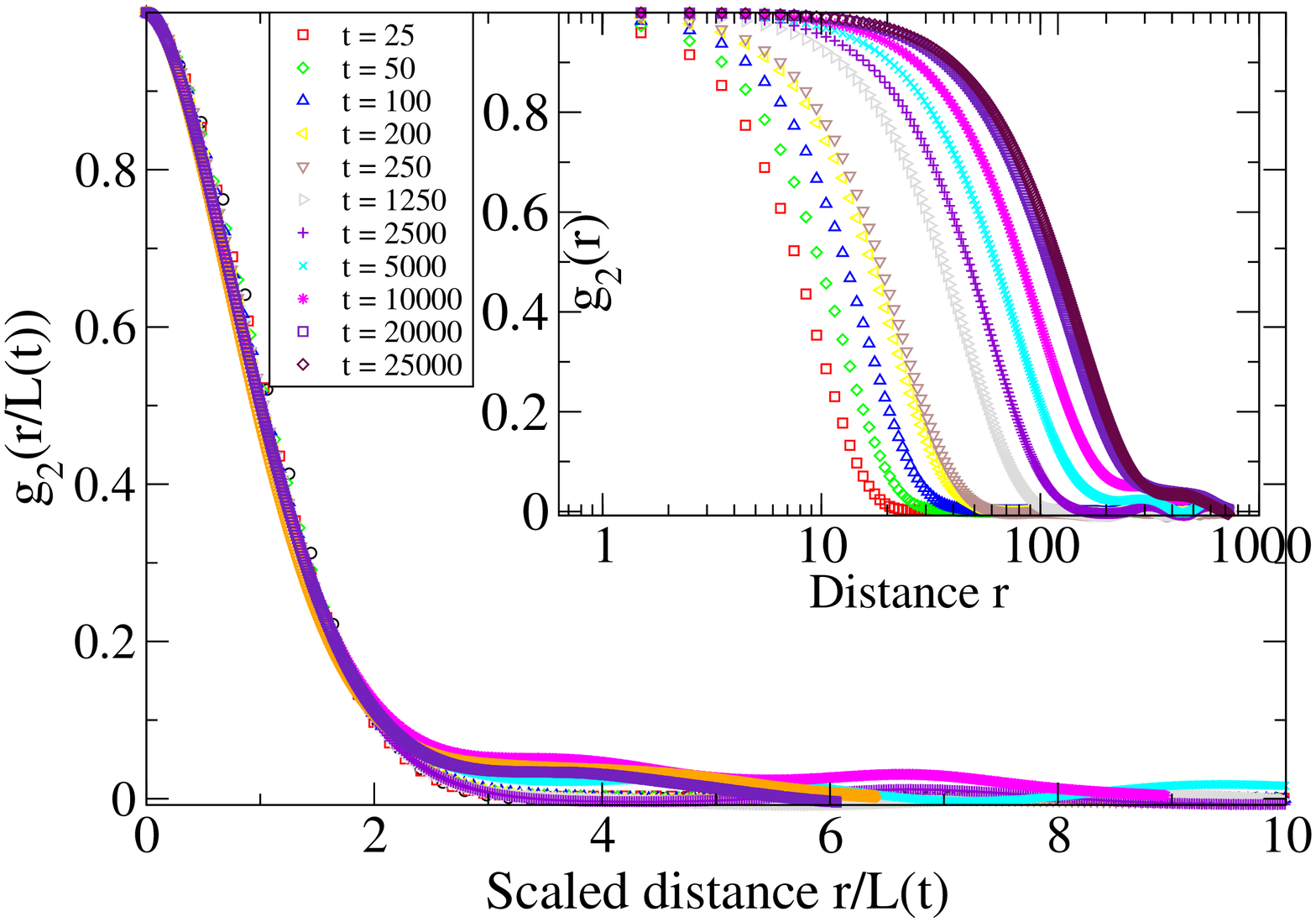}}
\subfigure(b){\includegraphics[height=6.0cm, width=6.0cm,
angle=270]{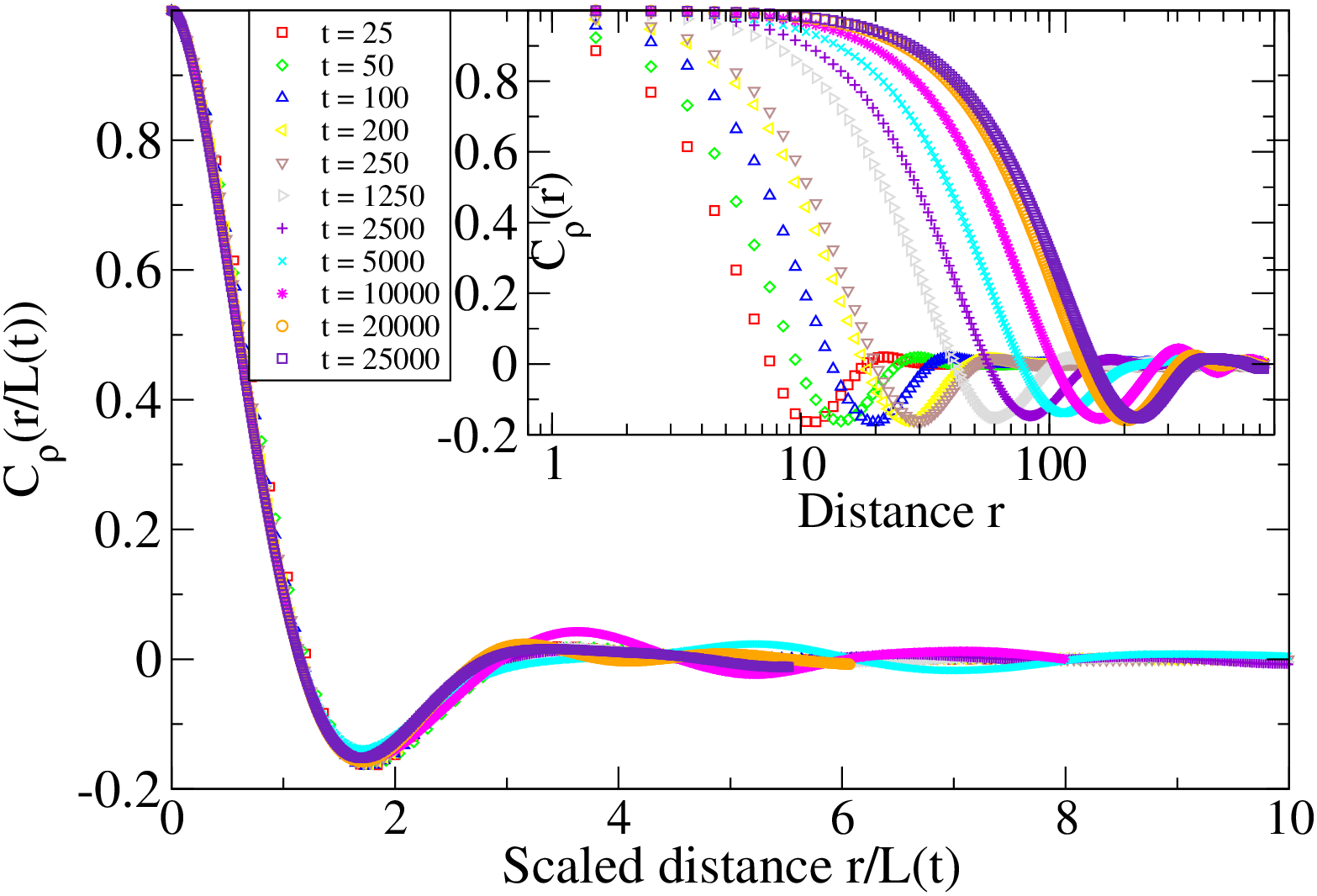}}
\caption{Evolution of the equal-time two-point nematic order parameter
correlator $g_2({\bf r}, t)$ (a) and density correlator $C({\bf r}, t)$ (b)
with time $t$, during domain growth, plotted vs.
$r/L(t)$, where $L(t)$ is the distance at which the correlator decays
Inset: correlators as functions of the unscaled $r$}
\label{corr_fns}
\end{center}
\end{figure}

In this section we present our numerical study of the growth kinetics of an
active nematic following a ``quench'' from an
isotropic initial state, as seen through \eqref{densityequation} and
\eqref{opequation},
with the parameter $\mu$ allowed to depend on the local density: $\mu(\rho) =
\rho - \rho_c$, where the threshold density $\rho_c = 0.5$ and the system is
prepared in a nearly uniform initial state with $\bsf{w} = 0$ and 
$\rho$ at each grid point drawn independently from a Gaussian distribution
with mean $\rho_0 = 0.75$ and standard deviation $0.05$. The cubic term in
(\ref{opequation}) is taken to have coefficient $2 \xi = 1$. The
system is thus in an unstable situation with respect to the growth of nematic
order, i.e.,
$\bsf{w}$. The value of $\mu(\rho_0)$ lies well beyond the banding instability
\cite{shi1,shi2,njop,ngo2014,putzig2014} in the regime where the long-time state
should be a homogeneous nematic. As we remarked in
the Introduction, we do not include the effects of noise in this study.
Randomness enters to a limited extent only via initial conditions. Such an
approach is adequate when
studying the domain growth of equilibrium phases, where it is established
\cite{abray} that the long-time, large-scale behaviour is dominated by a
zero-temperature fixed point. No comparable result is available for the
growth
kinetics of a nonequilibrium steady state like the active nematic, so our study
must be viewed as a first step. The role of the noise terms \cite{njop} in
coarsening is currently under investigation.

Following the quench, nematic order
appears locally, and the mismatch between the order in neighbouring regions
leads to the formation of elementary topological defects or disclinations
situated at points in the domain where the magnitude of the order parameter
vanishes. If one traces an anticlockwise loop enclosing any one such point,
the orientation traverses the range from $0$ to $\pm \pi$ for a
defect of strength $\pm 1/2$. These are seen in Fig. \ref{defectpic} where the
interval
$[0,\pi]$ is divided into five equal colour-coded sectors for visualisation.
With the passage of time defects of opposite strength find each other and
annihilate, leading to a well-ordered nematic. This is best seen through
measurements of the equal-time correlation function $g_2(r,t) = \mbox{Tr}\langle
\bsf{Q}({\bf 0},t) \bsf{Q}({\bf r},t) \rangle$ of the nematic order parameter
$\bsf{Q}$, as a function of spatial separation $r$ for different times $t$, as
presented in Fig. \ref{corr_fns}. Note: the angle brackets in the
definition of $g_2$ as well as that of the density correlator $C(r)$ below
imply an average over all pairs of points separated by a distance $r$ and over
ten initial confgurations. We define a characteristic length scale $L(t)$ as
that value of $r$ for which $g_2(r)$ decays to $0.1$ of its value at $r=0$, and
find that graphs of $g_2(r,t)$ \textit{vs} $r/L(t)$ for different $t$ collapse
onto a single curve; see Fig. \ref{corr_fns}(a).

\begin{figure}[htbp]
\begin{center}
\subfigure{\includegraphics[height=6cm, width=6cm,
angle=270]{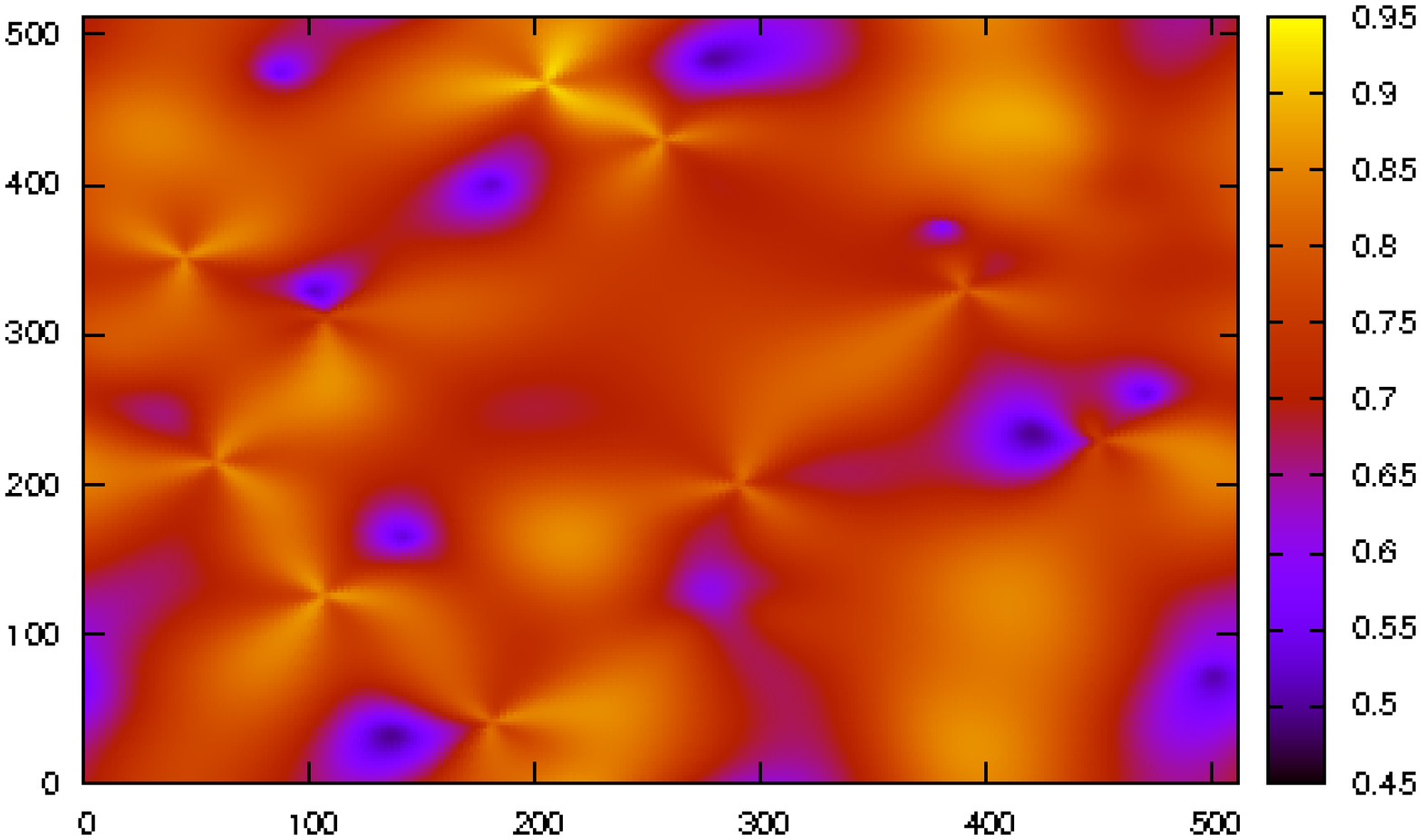}}
\subfigure{\includegraphics[height=6cm, width=6cm]{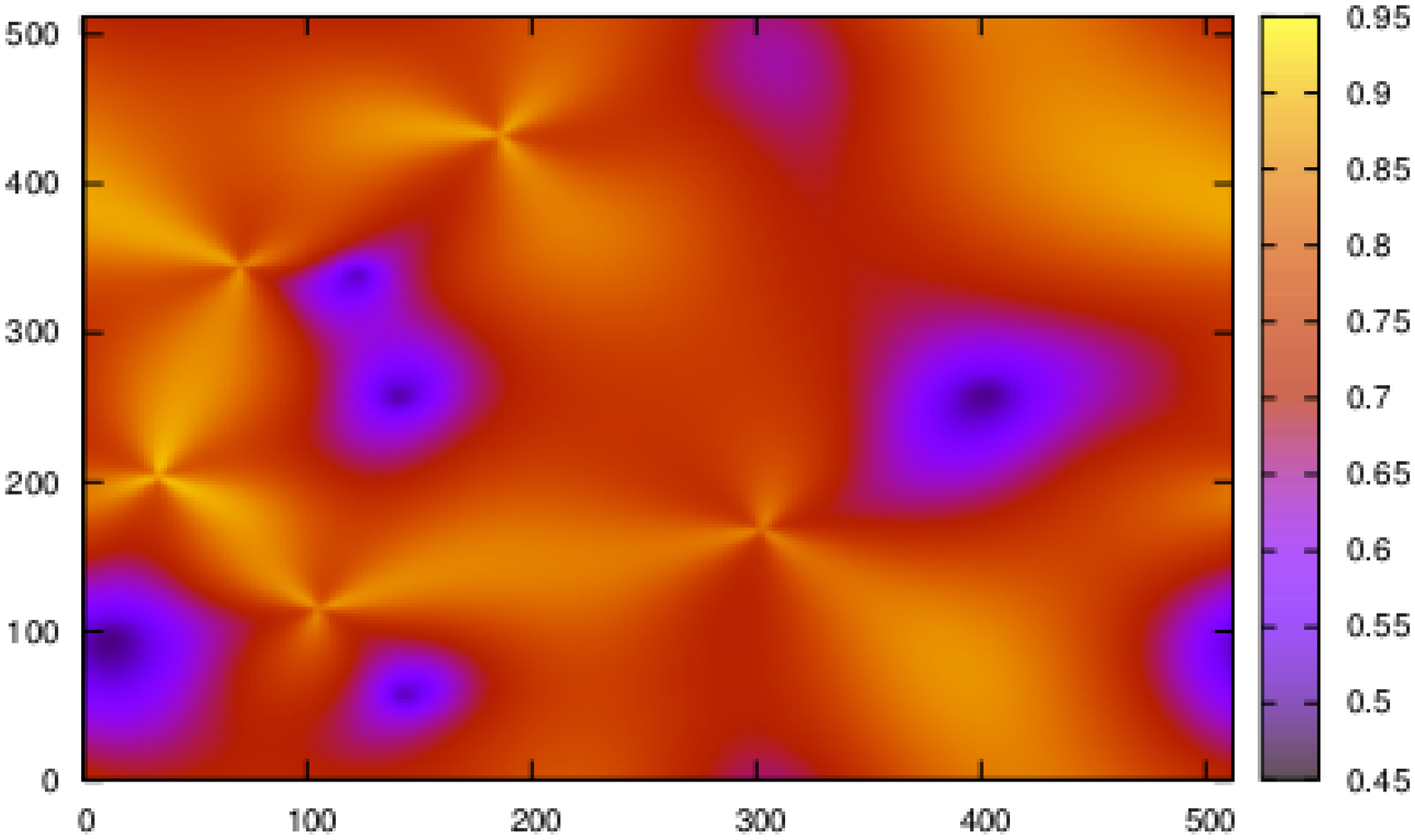}}
\caption{Density field during the growth of nematic
order, from numerical solutions of (\ref{densityequation}) and
(\ref{opequation}). Left and right figures show an earlier and a later time
configuration. Note the three-armed structures of high density,
and the large density variation overall. The order parameter, not shown
here, tracks the density -- dense regions have high order -- and the core
of the Y holds a strength $- 1/2$ type defect.}
\label{coarsenfig}
\end{center}
\end{figure}

We will examine below the domain growth law, i.e., the scaling of $L(t)$ with
$t$; let us look first at some interesting features of the morphology. Emanating
from each $-1/2$ defect is a characteristic Y shape, whose three arms are
regions with large values of the magnitude of the order parameter $\bsf{Q} =
\bsf{w}/\rho$ with one principal axis oriented along the arm. Neighbouring
regions outside the Y have much smaller values of $||\bsf{Q}||$. Moreover, and
this is our central result, the large-$||\bsf{Q}||$ regions are also
large-$\rho$ regions (see Fig. \ref{coarsenfig}). The contrast between low and
high density
regions is as much as a factor of two. Thus, the process of
coarsening of nematic order is accompanied by the onset of large inhomogeneities
in the density.

To examine further the nature and evolution of structure in the
density field, we measure the two-point density correlation function $C(r, t) =
\langle \delta \rho({\bf 0}, t) \delta \rho({\bf r}, t) \rangle$, where $\delta
\rho({\bf r},t) = \rho({\bf r},t) - \rho_0$ is the deviation of the local
density from the mean $\rho_0$. Like $g_2$ above, $C(r, t)$ also coarsens. Again
we define a length scale $L(t)$ as the value of $r$ at which the function
decreases to $0.1$ of its value at $r=0$. Fig. \ref{corr_fns}(b) shows a data
collapse when
$C(r, t)$ is plotted as a function of $r/L(t)$.

\begin{figure}[htbp]
\begin{center}
\subfigure{\includegraphics[height=6cm, width=6cm,
angle=270]{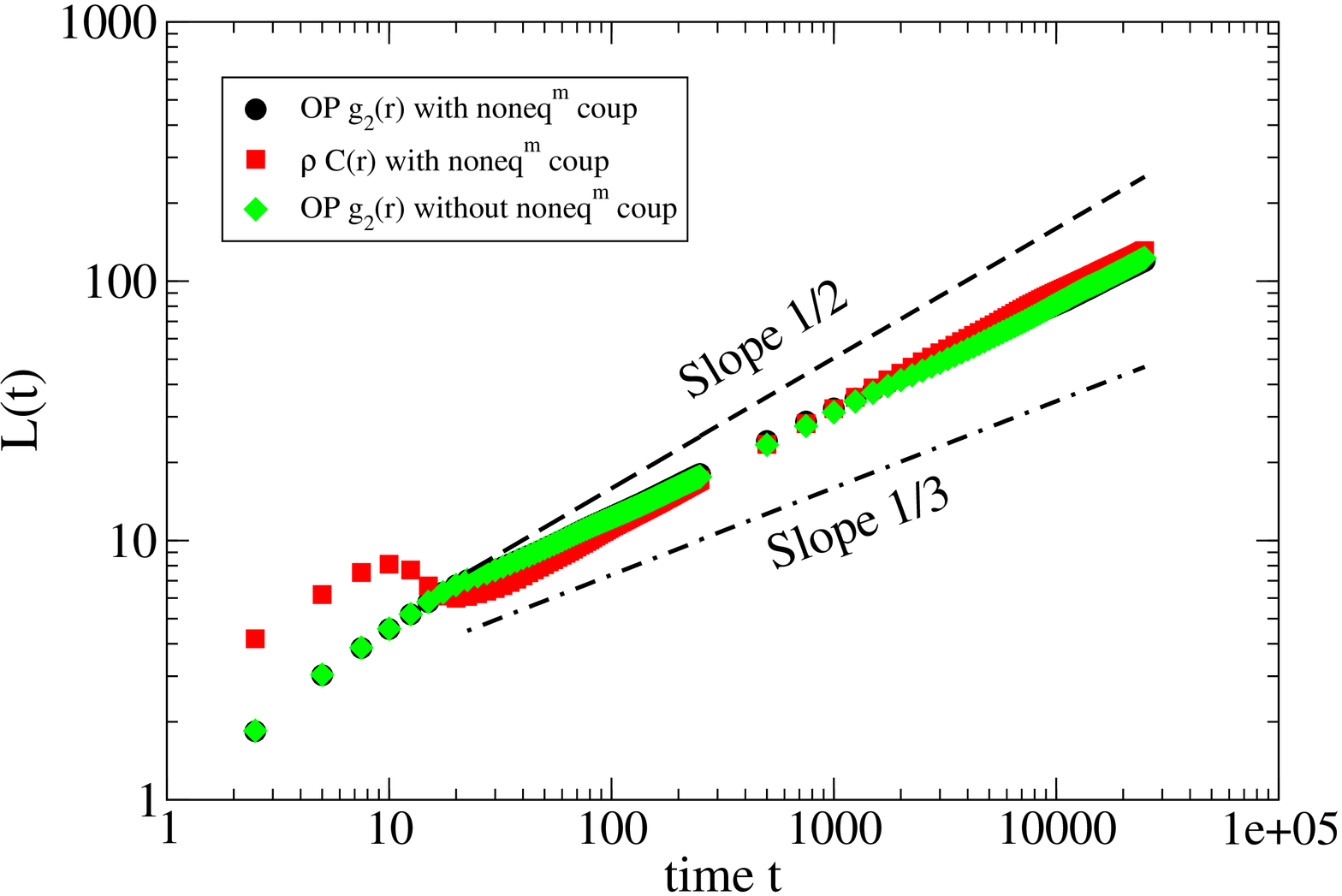}}
\subfigure{\includegraphics[height=6cm, width=6cm,
angle=270]{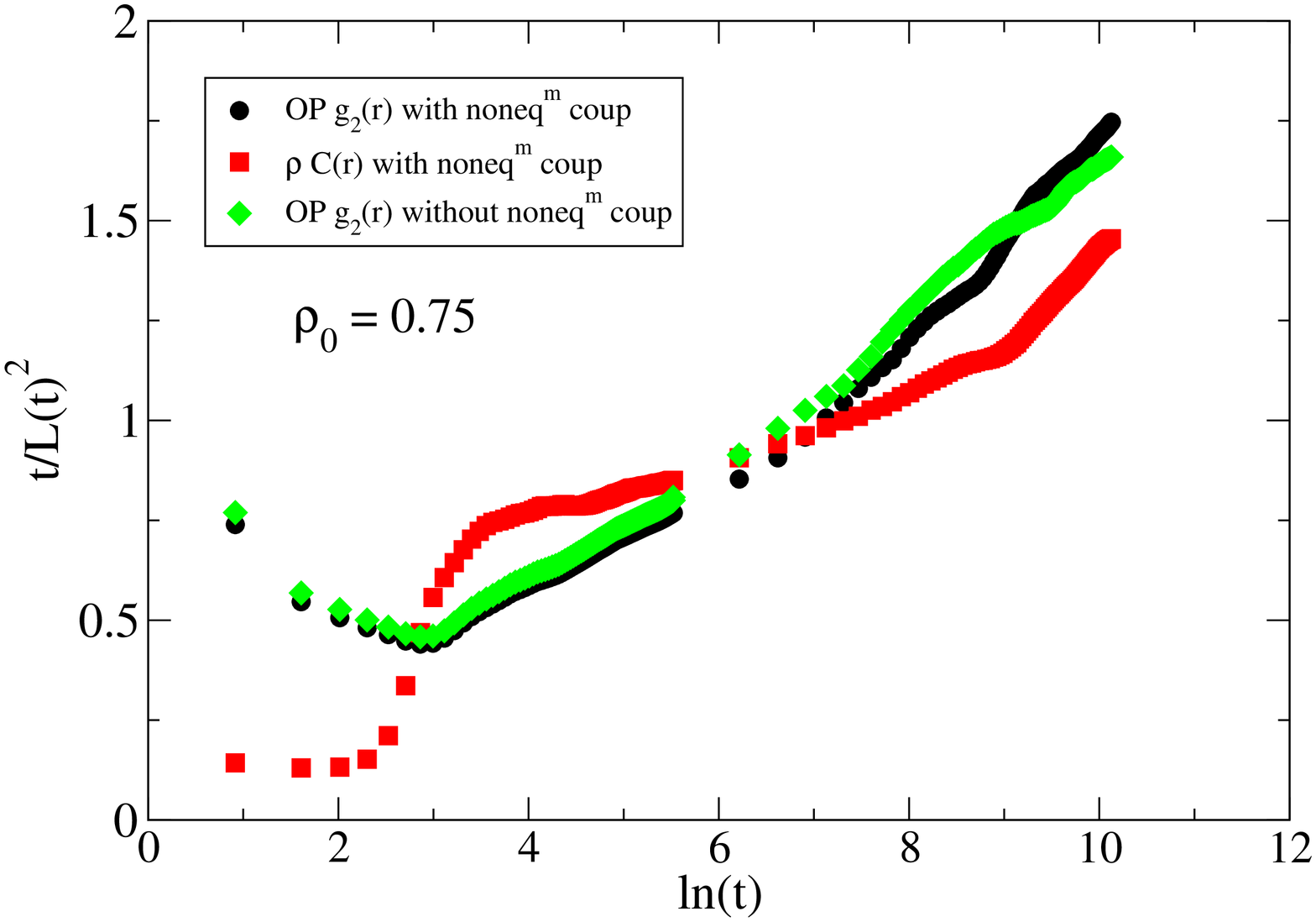}}
\subfigure{\includegraphics[height=6cm, width=6cm,
angle=270]{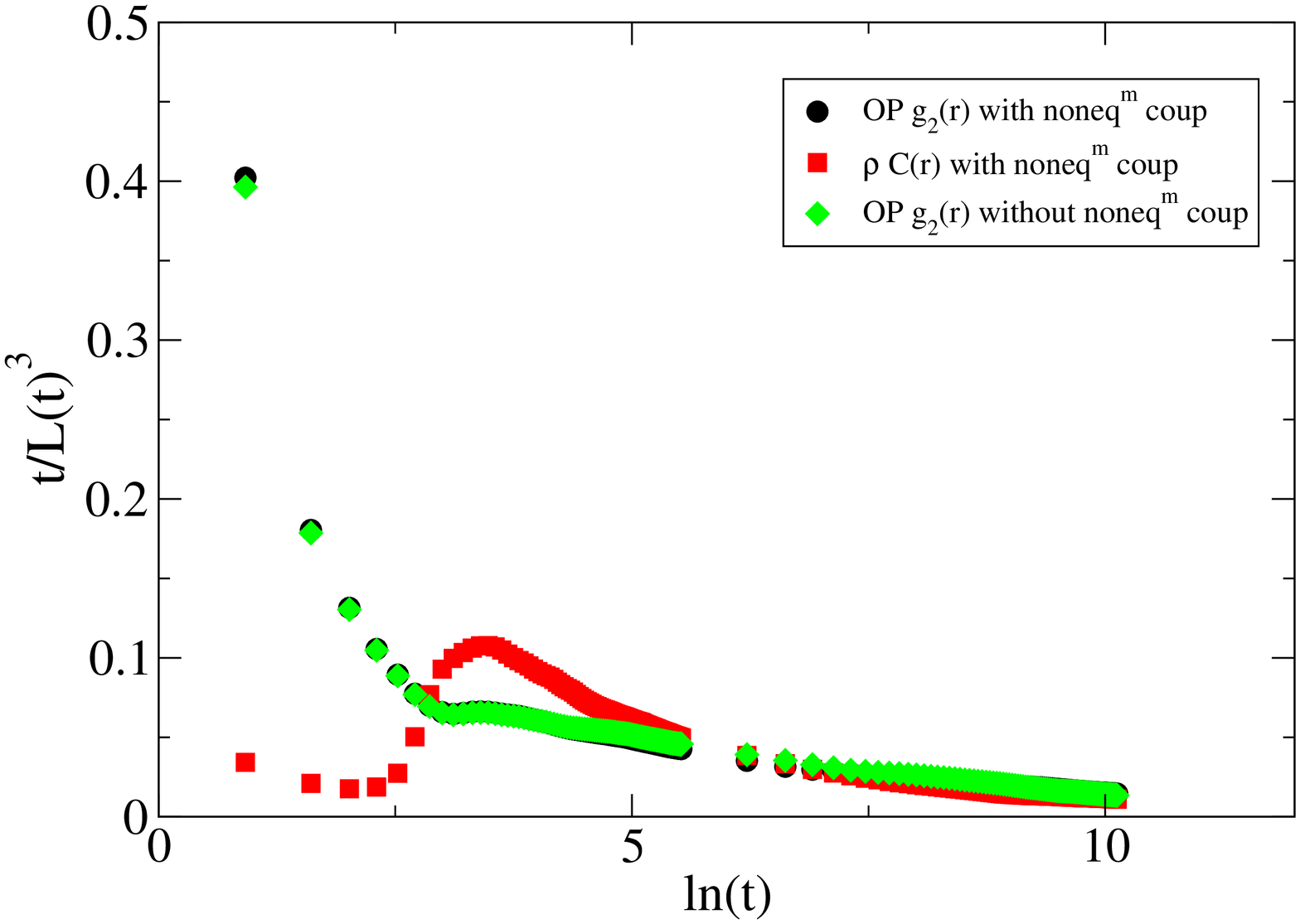}}
\caption{Upper left: characteristic length $L(t)$ from normalised two-point
correlators of order parameter and density vs. time $t$ on a log-log scale.
Straight lines are with slope 1/2 and 1/3 for comparison. Upper right: $t/L^2$
vs $\ln t$. Lower: $t/L^3$ decreases consistently with
time, suggesting the growth law is distinct from $t^{1/3}$.}
\label{lengthfig}
\end{center}
\end{figure}
\begin{figure}[htbp]
\begin{center}
\subfigure{\includegraphics[height=6cm, width=6cm,
angle=270]{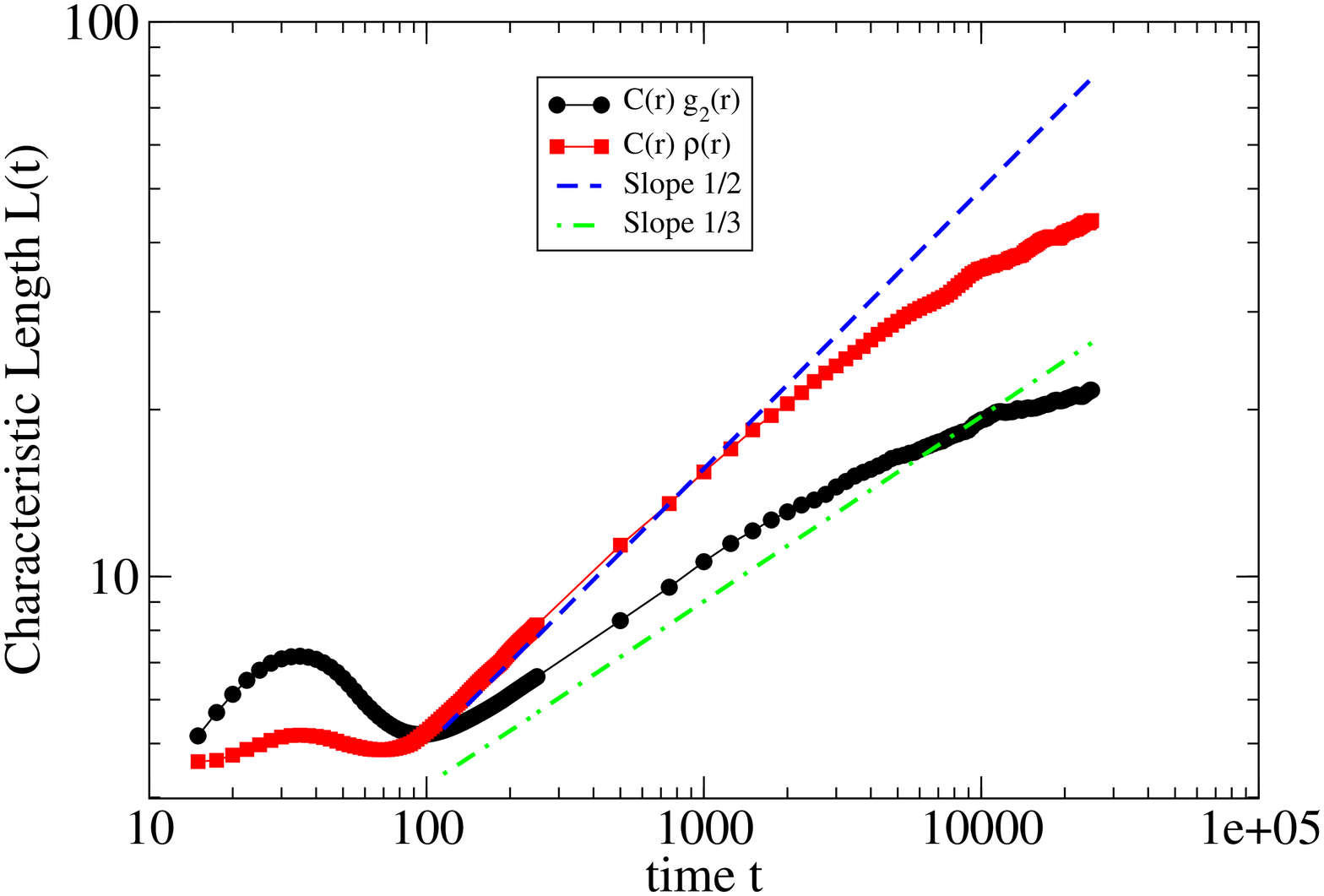}}
\subfigure{\includegraphics[height=6cm, width=6cm,
angle=270]{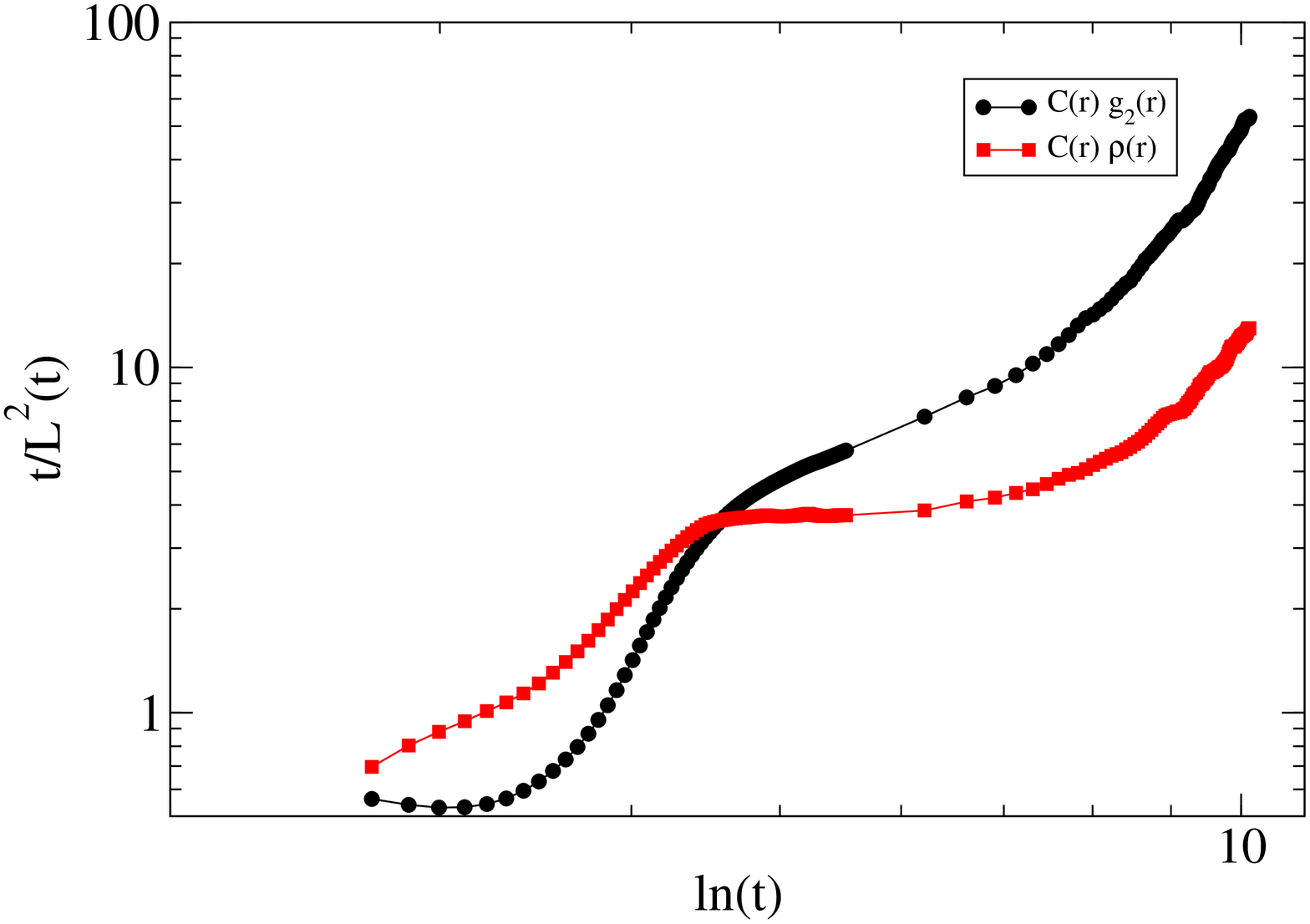}}
\subfigure{\includegraphics[height=6cm, width=6cm,
angle=270]{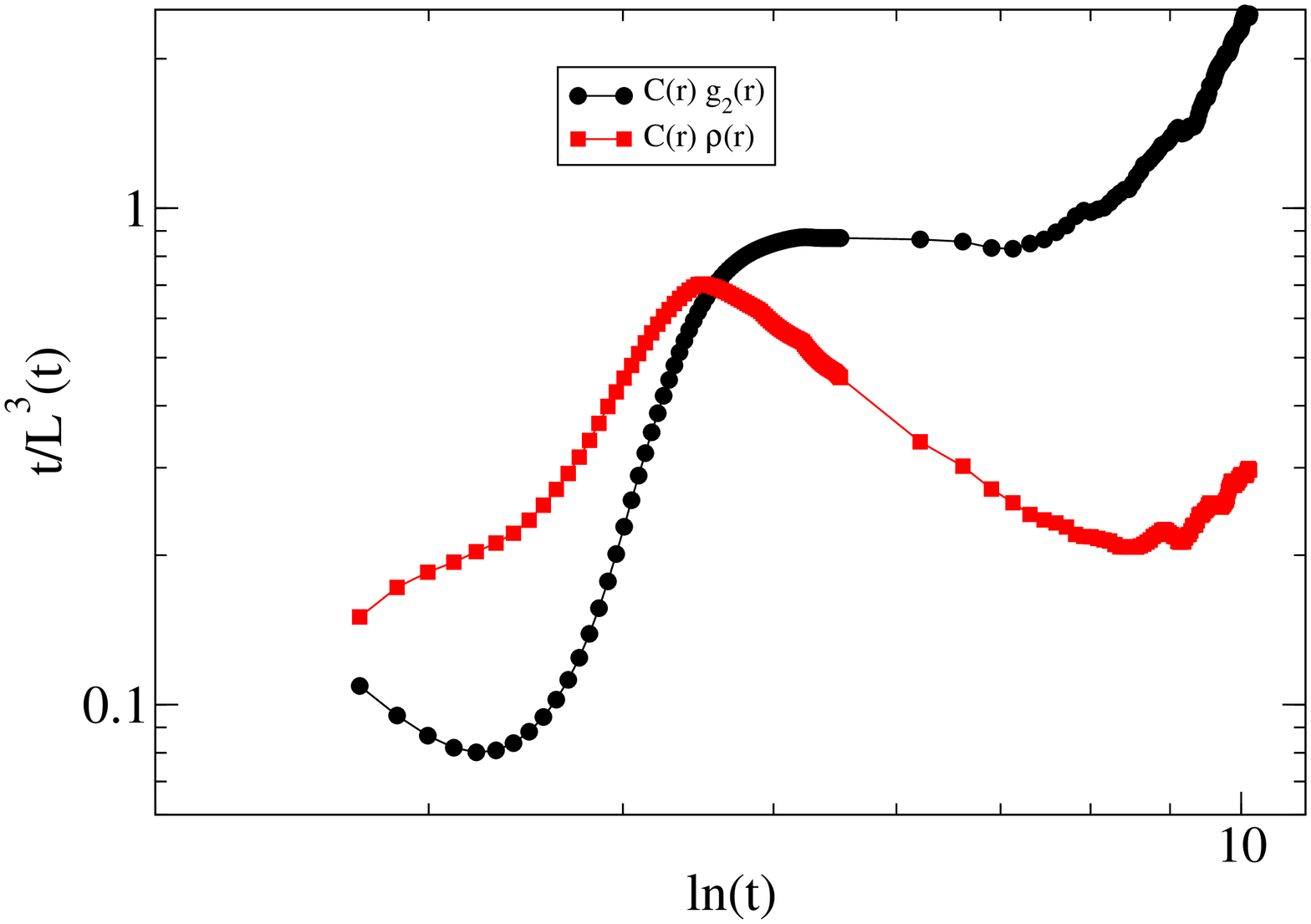}}
\caption{Upper left: characteristic length $L(t)$ from normalised two-point
correlators of order parameter and density vs. time $t$ on a log-log scale when
we are close to the isotropic-nematic transition, \textit{in the regime of the
linear instability to band formation}.
Straight lines of slope 1/2 and 1/3 are shown for comparison. Upper
and lower right: $t/L^2$
vs $\ln t$ and $t/L^3$ vs $\ln t$ respectively, suggesting that the density
coarsening is faster than $t^{1/3}$ and consistent with $t^{1/2}$, but that
nematic ordering is weak or arrested.}
\label{lengthfigunstable}
\end{center}
\end{figure}
Equations (\ref{densityequation}) and (\ref{opequation}), which are
used for the numerical
study, are known to possess a banding instability \cite{shi1} for parameter
values just past the mean-field onset of the nematic phase. To avoid this
instability,
we study these equations deep in the nematic phase, far from the transition
point. The lengths inferred from the coarsening of nematic order and the
clumping  of the density show essentially the same time dependence (Fig.
\ref{lengthfig}), consistent with a growth law with an exponent slightly smaller
than $1/2$ (but distinctly larger than $1/3$, which shows that $t/L^3$ decreases
steadily with increasing $t$). The {upper right frame} of Fig.
\ref{lengthfig} plots $t/L^2$ vs $\log t$; the roughly linear dependence is
consistent -- although not definitively so -- with $L \propto (t/\log t)^{1/2}$,
the law expected for the growth of an equilibrium nematic \cite{abray}. For
comparison, we also calculate the lengths for coarsening of the nematic order
parameter and density when we are in the banding regime (Fig.
\ref{lengthfigunstable}), and find the density still coarsens much as in Fig.
\ref{lengthfig} but the nematic order fails to keep up, consistent with the
approach to a phase-segregated but statistically isotropic
state \cite{shi1,ngo2014}.

The reason for the clumping is not far to seek. The nematic orientation
field has a three-fold symmetric structure around a $-1/2$ defect. The resulting
curvature, via the term ${1}/{2} {\bf \Gamma}:\bsf{w}$ in
(\ref{densityequation}), derived for the microscopic model
\cite{chateginellimontagne,njop} as outlined in section
\ref{sub:chatemodel}, drives currents in (\ref{densityequation}), towards
the defect core. This results in a Y-shaped density pattern. The ${\bf \Gamma}$
term in the order-parameter equation (\ref{opequation}) promotes anchoring of
the nematic orientation with respect to density gradients. The $\rho$ dependence
of $\mu$ in (\ref{opequation}) leads to loss and gain, respectively, of nematic
order in the resulting low- and high-density regions. Minor variations in the
microscopic model will lead to equations of motion of the same form but with
changes in the magnitude and, in principle, the signs of coefficients of the
${\bf \Gamma}$ term discussed above as well as other terms. 

Independent of the specific form of the coarsening laws, we emphasise that
clumping during the coarsening of an active nematic is a rather remarkable
finding. An initial state with uniform density condenses and coarsens
spontaneously. The evolution of the density is formally similar to that taking
place during phase separation, despite the absence of an attractive interaction
or a negative second virial coefficient. Note, in particular, that it is
distinct from the giant number fluctuations in a noisy active nematic steady
state, although the curvature-induced current term proportional to ${\bf
\Gamma}$ in (\ref{densityequation}) is responsible for both. What distinguishes
the process is that it takes place side-by-side with, and as a by-product of,
the coarsening of nematic order. Of course in our noise-free system, in a
situation where the uniform nematic is linearly stable at all wavenumbers, the
ultimate final state must be a quiescent, homogeneous nematic of uniform
density. It is important to note that the mechanism driving the clumping in our
case is distinct from the banding instability that arises just past the
mean-field threshold for the isotropic-nematic transition in an active system
\cite{shi1,njop,ngo2014}. We close this section by noting that a complete
description of the final state is incomplete without the inclusion of noise.
When noise of the appropriate type \cite{njop} is taken into account, we expect
interesting modifications to the correlation functions. We are currently
investigating the effect of noise on the correlators during coarsening
\cite{unpub}. At long times, these must cross over to the equal-time correlators
of the fluctuating active nematic steady state \cite{sradititoner,ngo2014}. 

\section{An orienting field suppresses density fluctuations}
\label{sec:extfield}
The study of \cite{sradititoner} says giant number fluctuations (GNF) are a
consequence of curvature-induced currents, i.e., that large-scale orientational
fluctuations are responsible for the currents that give rise to large density
inhomogeneities.
Since the  cause of GNFs in the steady state is long-wavelength fluctuations in
the angle field, we ask what happens if the latter are suppressed by an external
aligning agency, as can be accomplished in a conventional liquid-crystal system
by imposing a magnetic or electric field. Our findings offer a method to check
in principle whether the large density fluctuations in experiments
\cite{vjmenonsr} on active nematics are indeed a result of the mechanism of
\cite{sradititoner}. The idea is simple: if we apply an external field to a
nematic, the restoring force in response to an orientational distortion will
not vanish even in the limit of a spatially uniform change in direction. The
amplitude of orientational fluctuations will then saturate rather
than diverge at the smallest wavenumbers. The arguments in section
\ref{sub:curvcurr} then tell us that the density fluctuations will be cut off as
well. 

To test this idea, let us obtain a convenient form for the density
fluctuations in the presence of an orienting field. As the nematic order
parameter is a traceless symmetric tensor, such a field ${\bf h}$ imposed on a
equilibrium nematic will lead to a term $\bsf{Q}:{\bf h} {\bf h}$ in
the governing free-energy density. Note that we absorb details of
material parameters involved in such a coupling into a redefinition of ${\bf
h}$, and choose the sign of the coefficient to be positive without loss of
generality by exploiting the $\bsf{Q} \to - \bsf{Q}$ symmetry \cite{pgdg} of
two-dimensional nematics, the case of particular interest here. The result is
that the equation of
motion (\ref{opequation}) for the order parameter is modified by the presence
of a term $-{\bf h} {\bf h} + (h^2/2) \bsf{I}$ on the right-hand side,
favouring a mean state in which one eigendirection of
$\bsf{Q}$ is along ${\bf h}$. Perturbations about this mean state relax at a
rate that approaches a nonzero value of order $h^2$ in the limit of small
wavenumber. To assess the effect of this aligning field on density
fluctuations requires, strictly speaking the inclusion of a noise in the
equations of motion, with nontrivial dependence on the local values of $\bsf{Q}$
and $\rho$ as discussed in \cite{njop}. A complete treatment of the stochastic
PDEs for this problem, with a ``faithful'' rendition of the noise is a subject
of current study. In order to get a qualitative idea of how a field suppresses
density fluctuations, we introduce noise in a minimal manner, only in the
orientational dynamics. We represent the nematic order parameter in
terms of an amplitude $S$ and an angle $\theta$, $\bsf{Q} =
\left(\begin{array}{cc}
A & B\\ B & -A\end{array} \right)$ with $(A,B) =({S}/{2}) (\cos2\theta,\sin 2
\theta)$, and introduce a spatiotemporally white noise
$f_{\theta}({\bf r}, t)$ with correlator $\langle f_{\theta}({\bf 0}, 0)
f_{\theta}({\bf r}, t) \rangle
= 2\Delta_{0} \delta({\bf r})\delta(t)$. We work with $\Delta_0 = 0.03$, and
an aligning field ${\bf h} = (h,0)$ with $h^2 =  0.001, 0.005, 0.01, 0.05$
and $0.1$. {In general, the hydrodynamic equations of motion
derived from the microscopic rule have multiplicative noise terms \cite{njop,shradhathesis}. 
However, for simplicity, we use additive spatiotemporally white noise.} 
We study the resulting stochastic partial differential equations
numerically in the same regime deep in the nematic phase as was explored in the
coarsening studies. We compare the measured density correlation functions with
approximate analytical expressions from a linearised treatment of small
fluctuations about the mean uniform ordered phase. We focus on the equal-time
spatial autocorrelator $S_{\bf q}$ of the density, as a function of wave-vector
${\bf q}$, at fixed mean density $\rho_0 = 0.75$. We present results only for
$q_x = q_z$. From the equations of motion presented above, a straightforward
calculation within a linearised approximation {(presented in Appendix A)}
will show that, in the presence of a field $h$, the standard deviation $\Delta N$ of
density fluctuations for a region with $N$ particles on average satisfies 
\beq
\label{Sqfield}
{(\Delta N)^2 \over N}  = \mathcal{K} + \frac{N}{(a+N h^2)^2}
\eeq
\begin{figure}[htbp]
\begin{center}
{\includegraphics[height=9cm, width=10cm,
angle=270]{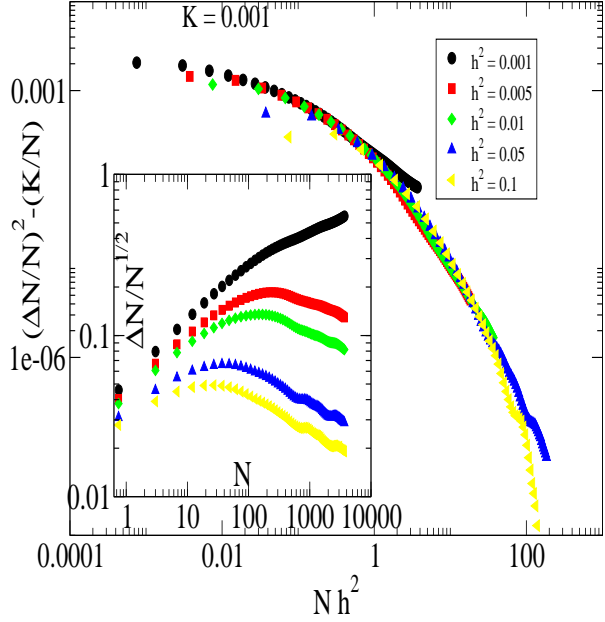}}
\caption{(color online) An orienting field limits giant number
fluctuations: scaled and background-subtracted variance, for different values
of field, plotted against scaled mean number, lie on a single curve. Inset:
unscaled data, showing the suppression of number fluctuations at large scales
in the presence of a field.}
\label{numfluc_ext}
\end{center}
\end{figure}
where $\mathcal{K}$ is a finite background coming from local number-conserving
processes and $a$ is a constant. Even in the absence of a noise
directly driving the $\rho$ equation, nonlinear effects in
(\ref{densityequation}) and (\ref{opequation}) are expected to lead to such a
background, so we retain it when fitting our data. Note that the second term in
(\ref{Sqfield}) contains the giant number fluctuations \cite{sradititoner}
modified by the field $h$. The form of that term implies that as one looks
at regions of increasing size and hence increasing $N$, the excess relative
density fluctuations initially grow as they would without a field and then
decrease once $N$ goes past a crossover value of order $h^{-2}$. We compare
(\ref{Sqfield}), viewed as an expression for the density structure factor
at wavenumber $q \sim N^{-1/2}$, to numerical solutions of
(\ref{densityequation}) and (\ref{opequation}) with noise and field. We infer
a value of $\mathcal{K}$ by looking at the smallest wavenumber data in the
presence of a large field $h$, and subtract that value uniformly from all our
data. The resulting quantity, divided by $N$, should yield a data collapse when
plotted against $Nh^2$. Reasonably satisfactory results along these lines can
be seen in Fig. \ref{numfluc_ext}, with a background subtraction of $\mathcal{K}
= 0.001$.
We emphasise that our aim, from these studies with limited dynamic range, is to
underline the role of an externally imposed field as a way of testing the origin
of the giant density fluctuations, rather than arguing for a particular value of
scaling exponents (see \cite{ngo2014}).

\section{Other issues} \label{sec:otherissues}
\subsection{Correlation vs response}\label{corr_resp}
At an {\em equilibrium} critical point, density fluctuations are large
essentially because the response function is large and slowly decaying. This
means that if the density at a point is held at a value different from the
equilibrium value, there will be a slowly decaying density profile around it. In
an active nematic, as can be seen in \cite{sradititoner} the fluctuations are
large not because of the density propagator, but because orientational
fluctuations enter the density dynamics as correlated, nonequilibrium noise. A
formal calculation of the response of the density to a change in chemical
potential will reveal nothing singular. Is there, then, a field to which the
mean density response is large and slowly decaying? The answer is yes, the
orientation. Thus, a field perturbing the orientation field will produce
long-range distortions in density. In addition, suppressing spontaneous
fluctuations of the nematic, by an external field, should kill the giant number
fluctuations, as we showed in section \ref{sec:extfield}. 

\subsection{The meaning of anisotropic large-scale correlations in an
active nematic}
\label{subsec:aniso}
In two space dimensions, the setting for most studies referred to in this paper,
particle-based simulations \cite{chateginellimontagne,ngo2014} (as well as a
renormalisation-group (RG) treatment of nonlinearities in the limit where the
density field is ignored \cite{rgshradhasriram}) find that the active nematic
phase is distinguished from the isotropic phase only by the presence of
power-law orientational correlations; the order parameter goes to zero
algebraically with increasing system size. Concomitantly, the nonequilibrium
analogues to the two Frank elastic constants for a two-dimensional (2D) active
nematic flow to equality at large length scales under the RG as they do in an
equilibrium nematic \cite{pelc}. Although
quasi-long-range order is standard in equilibrium 2D nematics, the active
current (\ref{jact}) \cite{sradititoner} is explicitly anisotropic
when expanded about the ordered nematic phase, with $x$ and $z$ components
respectively proportional to $\partial_z \theta$ and $\partial_x \theta$ for
slowly-varying angular fluctuations $\theta$. The predicted
consequences are \textit{anisotropic} power-law density correlations at small
wavevectors ${\bf q} = (q_x,q_z)$, of the form $q_x^2q_z^2/q^6$.
On the other hand, quasi-long-range order, as known to hold in two-dimensional
active  nematics as well, implies the absence of a true nematic on the largest
length scales, and thus allows no global anisotropy even in the sense of a
spontaneously broken symmetry. Thus, just as nonlinearities must cause the
system to flow to a one-Frank-constant picture at large scales
\cite{rgshradhasriram} when the density is ignored, they presumably also
isotropise the density correlator. From ref. \cite{ngo2014}, it would appear
that they also change its scaling from anisotropic $1/q^2$ to isotropic
$1/q^{2-\eta}$. A renormalisation-group theory of the value of $\eta$ remains a
challenge. 

\begin{acknowledgments} We thank H Chat\'e and F Ginelli for useful
discussions, SM acknowledges support from a DST INSPIRE Faculty Fellowship,
and SP \& SR were supported in part by a J.C. Bose Fellowship.
\end{acknowledgments}



\newpage
\section*{Appendix A: Active nematic with an external field}
Using the representation 
$ \Q  = \left(\begin{array}{cc}
A & B\\
B & -A\end{array} \right)$, 
where $ (A, B) =\frac{S}{2} (\cos2\theta, \sin2\theta)$, 
taking the density $\rho = \rho_{0} + \delta \rho$, 
where $\rho_{0}$ is the mean density, keeping terms to  linear order in $\theta$ and  $\delta \rho$, and treating $S$ as constant we get
\begin{equation}
\frac{\partial \delta \rho}{\partial t} = B_{1}\partial_{x}^2 \delta \rho + B_{2}\partial_{z}^2 \delta \rho + D_{2}\partial_{x}\partial_{z} \theta  
\label{appendix1eq}
\end{equation} 
\begin{equation}
\frac{\partial \theta}{\partial t} = A_{1}\partial_{x}^2\theta + A_{2}\partial_{z}^2\theta + D_{1}\partial_{x}\partial_{z} \delta \rho - h^2\theta +f_{\theta}
\label{appendix2eq}
\end{equation}
where $h$ is the strength of the external applied field. 
$f_{\theta}({\bf r}, t)$ 
is non-conserving spatiotemporal Gaussian random white noise 
with noise-noise correlation 
$<f_{\theta}({\bf r}, t) f_{\theta}({\bf r}', t') > 
= 2\Delta_{0} \delta({\bf r}-{\bf r}')\delta(t-t')$. 
$\Delta_{0}$  
is noise strength for orientation equation. 
Writing equations ({\ref{appendix1eq}} and {\ref{appendix2eq}}) in Fourier space we get
\begin{equation}
(-i\omega+A_{1}q_{x}^2+A_{2}q_{z}^2 + h^2)\theta({\bf q}, \omega) = -D_{1}q_{x}q_{z} \delta \rho({\bf q}, \omega) + f_{\theta}({\bf q}, \omega) 
\label{appendix3eq}
\end{equation}
\begin{equation}
(-i\omega+B_{1}q_{x}^2+B_{2}q_{z}^2)\delta \rho({\bf q}, \omega) = -D_{2}q_{x}q_{z} \theta({\bf q}, \omega) -i {\bf q} \cdot {\bf f}_{\rho}({\bf q}, \omega). 
\label{appendix4eq}
\end{equation}
Solving  these two coupled linear equations 
({\ref{appendix3eq}, \ref{appendix4eq}}) for 
$\theta({\bf q}, \omega)$ and $\rho({\bf q}, \omega)$, we see
\begin{align}
&\left(\begin{array}{cc}
-i \omega +A_{1}q_{x}^2 +A_{2} q_{z}^2 + h^2 & D_{1}q_{x} q_{z}  \\
D_{2}q_{x} q_{z} & -i \omega +B_{1}q_{x}^2 +B_{2} q_{z}^2\end{array} \right)\left(\begin{array}{cc} \theta({\bf q}, \omega) \\
\delta \rho({\bf q}, \omega) \end{array} \right) = \left(\begin{array}{cc} f_{\theta}({\bf q}, \omega) \\
-i {\bf q} \cdot {\bf f}_{\rho}({\bf q}, \omega) \end{array} \right)  \notag \\
& \implies \left(\begin{array}{cc} \theta({\bf q}, \omega) \\
\delta \rho({\bf q}, \omega) \end{array} \right) = {\bf M}^{-1} \left(\begin{array}{cc} f_{\theta}({\bf q}, \omega) \\
-i {\bf q} \cdot {\bf f}_{\rho}({\bf q}, \omega) \end{array} \right) 
\label{appendix5eq}
\end{align}
where ${\bf M} = \left(\begin{array}{cc}
-i \omega +A_{1}q_{x}^2 +A_{2} q_{z}^2 + h^2 & D_{1}q_{x} q_{z} \\
D_{2}q_{x} q_{z} & -i \omega +B_{1}q_{x}^2 +B_{2} q_{z}^2\end{array} \right)$ \\
and ${\bf M}^{-1} = \frac{1}{det[{\bf M}]}\left(\begin{array}{cc}
-i \omega +B_{1}q_{x}^2 +B_{2} q_{z}^2 & -D_{1}q_{x} q_{z} \\
-D_{2}q_{x} q_{z} & -i \omega +A_{1}q_{x}^2 +A_{2} q_{z}^2 + h^2\end{array} \right)$ 
Now the quantity in which we are interested  is the 
two-point density correlator, 
given by 
\begin{align}
<\delta \rho({\bf q}, \omega) \delta \rho(-{\bf q}, -\omega)> & 
= <|\delta \rho({\bf q}, \omega)|^{2}>  \notag \\
&= \frac{2(\omega^2+(A({\bf \hat q}) q^2 +h^2)^2)\Delta_{\rho}({\bf \hat q}) q^2+ 2D_{1}^2 q_{x}^2q_{z}^2 \Delta_{0}}{det[{\bf M}] det[{\bf M}^{*}]}
\label{appendix6eq}
\end{align}
where $det[{\bf M}] det[{\bf M}^{*}] 
= \omega^2((A({\bf \hat q}) + B({\bf \hat q})) q^2+ h^2)^2 
+ ((A({\bf \hat q})q^2+h^2)B({\bf \hat q}) q^2 
- D_{1}D_{2} q_{x}^2 q_{z}^2 -\omega^2)^2$. 
In the steady state one can calculate
\begin{equation}
S({\bf q}, t) = <\delta \rho({\bf q}, t) \delta \rho(-{\bf q}, t)> = \frac{1}{2\pi}\int_{-\infty}^{+\infty} <|\delta \rho({\bf q}, \omega)|^2> d\omega
\label{appendix7eq}
\end{equation}
which after integration over $\omega$ and for $q_x = q_z$ gives

\begin{equation}
\label{appendix8eq}
{(\Delta N)^2 \over N}  =  \frac{N}{(a+N h^2)^2}
\end{equation}
where $\mathcal{K}$ is a finite background coming from local number-conserving
processes and 
Where $a$ is a constant. In the absence of a noise
directly driving the $\rho$ equation, nonlinear effects in
(\ref{densityequation}) and (\ref{opequation}) are expected to lead  a
background, expression for ${(\Delta N)^2 \over N}$ is
\begin{equation}
\label{appendix9eq}
{(\Delta N)^2 \over N}  =  \mathcal{K} + \frac{N}{(a+N h^2)^2}
\end{equation}
so we retain it when fitting our data.


\begin{thebibliography}{10}
\bibitem{foot1noise} The equations are derived from a microscopic stochastic
theory and contain parameters whose value depends on noise strength at the
particle scale, but the coarse-grained equations are treated as describing the
average macroscopic behaviour and are hence non-stochastic.
\end{thebibliography}
\end{document}